# Cryogenic cleaning of tin-drop contamination on surfaces relevant for extreme ultraviolet light collection

## Norbert Böwering [1,2,a)] and Christian Meier [1]

[1] Molecular and Surface Physics, Bielefeld University, 33615 Bielefeld, Germany,
[2] BökoTech, Ringstr. 21, 33619 Bielefeld, Germany
[a)] Electronic mail: boewering@physik.uni-bielefeld.de

## Abstract

Improvement of tool reliability and uptime is a current focus in development of extreme ultraviolet lithography. The lifetime of collection mirrors for extreme ultraviolet light in tin-based plasma light sources is limited considerably by contamination with thick tin deposits that cannot be removed sufficiently fast by plasma etching. For tin droplet splats sticking to large substrates, we have developed and compared several efficient cleaning techniques based on cryogenic cooling. A silicon carbide substrate and different silicon wafer samples with up to 6 inch diameter with the surface uncoated, multilayer-coated, unstructured and grating-structured were tested. After tin dripping onto heated samples, embrittlement of droplet contamination is induced in-situ by stresses during phase transformation, following the initiation of tin pest with seed crystals of gray tin. Conversion of initially adhesive deposits to loose gray tin has been reached in less than 24 hours on all tested surfaces by continuous cooling with cold nitrogen vapor to temperatures in the range of -30 – -50 °C. Alternatively, stress-initiated tin-removal by delamination of β-Sn droplet splats has been attained via contraction strain induced by strong cooling to temperatures of around -120 °C. Profilometry has been used to analyze the bottom side of tin droplet splats removed from a grating-structured wafer. The in-situ tin cleaning techniques give results comparable to fast ex-situ cleaning that has been achieved either by sample immersion in liquid nitrogen or by splat removal after $CO_2$ snowflake aerosol impact using a hand-held jet-nozzle. The implementation of the in-situ phase-conversion concept for the cleaning of collector mirrors in commercial light sources for lithography is discussed.

# I. INTRODUCTION

After many years of development, optical projection lithography using scanners at extreme ultraviolet (EUV) wavelengths of around 13.5 nm has finally been adopted by the semiconductor industry for the production of critical layers in chip manufacturing.[1] The EUV light source created by the laser-produced plasma (LPP) with tin micro-droplet targets and light collection with multilayer coated mirrors (MLM) has now reached reliable performance levels at in-band EUV output powers of above 250 W and sufficiently high wafer throughput exceeding 125 wafers per hour.[2] As a next step, technology suppliers and users are concentrating on further increasing scanner reliability, uptime and tool productivity. Particularly critical parameters influencing the light source availability are the uptime of drive laser as well as the lifetimes of droplet generator and collector mirror. Substantial improvements have already been achieved for the performance of these crucial components in recent years.[2,3]

Nevertheless, since the collector mirror is located very close to the plasma, its reflectance still deteriorates during operation due to deposition of tin debris, in spite of several tin mitigation schemes that have been implemented during light source development by the suppliers.[3,4] The MLM reflectance of the collector becomes vanishingly small at locations covered by tin deposits with a thickness exceeding just a few nanometers. Besides primary debris originating directly from the droplet target, tin fragments and particles can impact the mirror as secondary debris after collisions or accumulation at vessel walls. Deposition on the mirror surface may occur by incident tin vapor, by micro- and macroscopic droplets, and even by millimeter-size drops of liquid tin that melt on the chamber walls when heated by the infrared drive laser. Gravity can cause tin drops to drip onto the collector mirror that is located near the bottom of the source chamber with its surface facing up.[4] The contamination of this mirror by tin drops can therefore be very severe, as was also documented with photos.[5]

In order to reduce tin deposition, internal cleaning may be applied by etching of tin via hydrogen radicals that are generated by the interaction of EUV radiation with the hydrogen buffer gas present in the source vessel. This leads to a slow removal of tin deposits through the formation of volatile tin tetra-hydride molecules that can be pumped away.[6-9] The tin removal rate could be increased further by additional generation of hydrogen radicals using capacitively-coupled or microwave-generated plasma arrangements.[10,11] However, such etching schemes are not sufficiently rapid for cleaning layers with a thickness exceeding several tens of micrometers in reasonably short times. When during source operation the reflectance of the mirror has dropped below a critical value, the collector module has to be swapped and cleaned externally. Dependent on the severity of contamination and damage, different methods may be chosen for ex-situ cleaning. Dry cleaning using gaseous or frozen-particle flows like carbon dioxide ($CO_2$) dry-ice pellets may be used[12], as was described for EUV light source optics and mask cleaning[13-16]. Chemical wet cleaning without or with partial coating layer removal and even more extensive surface refurbishment can also be applied.[17,18] However, in order to avoid lengthy collector swaps and related vessel pump-down periods causing prolonged source down-time, it is of high interest to investigate also alternative cleaning techniques for rapid removal of thick tin deposits that may be applied inside of the source vessel without requiring time-consuming mirror exchanges.

Previously, one of the authors has reported on a cleaning scheme based on the induction of tin pest by seed particles[19] where solidified drops of high-purity tin break up by embrittlement during phase transformation of β-Sn to the allotrope α-Sn.[20] This concept was also independently proposed for collector cleaning in two, meanwhile abandoned patent applications where, however, the important nucleation initiation process was not considered.[21,22] Recently, we have described first results for in-situ cleaning under high vacuum of tin drop deposits on small samples of silicon (Si) wafers with different multilayer coatings. Adhesion of tin splats to smooth surfaces was generally observed during dripping when the temperature of the samples was distinctly above room temperature. After contact with α-Sn seed-particles, phase transformation to gray tin was induced during substrate cooling to temperatures in a range between -25 °C and -40 °C.[23, 24] The recovery of EUV reflectance after removal of detached converted tin pieces was reported for an MLM sample with Mo/Si coating.[23] The temperature dependence of droplet sticking to the substrates was studied for different sample types.[24]

During the previous in-situ tin phase-transformation experiments we have examined small samples with smooth silicon substrates. In order to assess the cleaning of samples directly relevant for EUV radiation collection in commercial light sources we have now implemented the capability to study and compare large samples (up to 6 inches in diameter). Furthermore, a grating-structured sample surface and also a substrate made of silicon carbide (SiC) have been investigated. This was examined since collector mirrors with concentric grating structures on their surface are used in EUV sources to reduce the propagation of infrared laser radiation to the scanner by diffraction.[25,26] In addition, we compare in this study the method and efficiency of tin drop removal by embrittlement with several other cryogenic cleaning techniques tested and carried out both in-situ and ex-situ.

## II.  METHODS

To a large extend our setup for in-situ tin dripping and cleaning experiments has been described in detail previously.[23, 24] Briefly, it consists of an ultra-high vacuum chamber with several viewports and ionization gauge as well as quadrupole mass analyzer (Balzers QMA 150 with cross-beam ion source) for vacuum monitoring (base pressure without bake-out: $10^{-5}$ Pa). Flanges were sealed with either copper or viton gaskets. The chamber has a region for droplet preparation by melting of tin pieces in its upper section and a sample holder with cooling and heating capability in its lower section. Optical imaging of tin splats on the samples was made from the top of the chamber through a vacuum window using a digital camera with zoom objective (Canon EOS 350D) and illumination from a panel consisting of light emitting diodes and front diffuser.

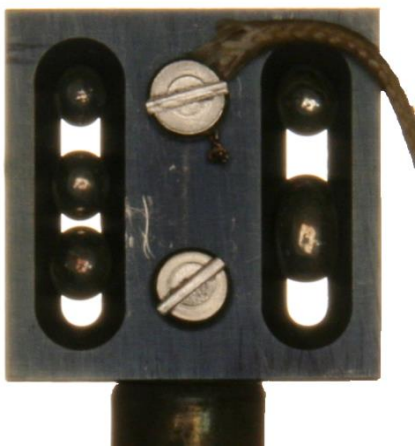

FIG. 1. Photo of heatable drip plate with 5 tin pieces on drip slots.

Before pump-down several pieces of tin (99.999% purity) with mass each in the range of 100 - 200 mg were placed on an insertable Mo drip tray with two 2 mm-wide slots. After evacuation the tray was resistively heated beyond the melting point of tin (232 °C) to temperatures $T_d$ of typically 250 – 300 °C. A thermocouple wire was used to monitor the temperature of the tray. Fig. 1 shows a photo of the Mo drip tray after the tin pieces had contracted to round balls during heating to above the melting point of tin. After reaching temperatures above ~250 °C the balls of molten tin dripped through the slots when the adhesion force to the tray was exceeded. When required for release of adhering drops, small vibrations could be induced in addition by tapping the tray support with a bar mounted on a rotation feedthrough.

At a distance of 45 cm below the tray, spherical tin drops with diameters D in the range of typically 3.0 – 3.6 mm hit the samples mounted in the lower section of the vacuum chamber after a fall of 0.3 s with an impact velocity of ~3 m/s and at Weber numbers of around 300 – 400. This resulted in the creation of circular solidified tin splats on the sample surface with diameters in the range of 8 – 11 mm. The dripping arrangement is illustrated schematically in Fig. 2.

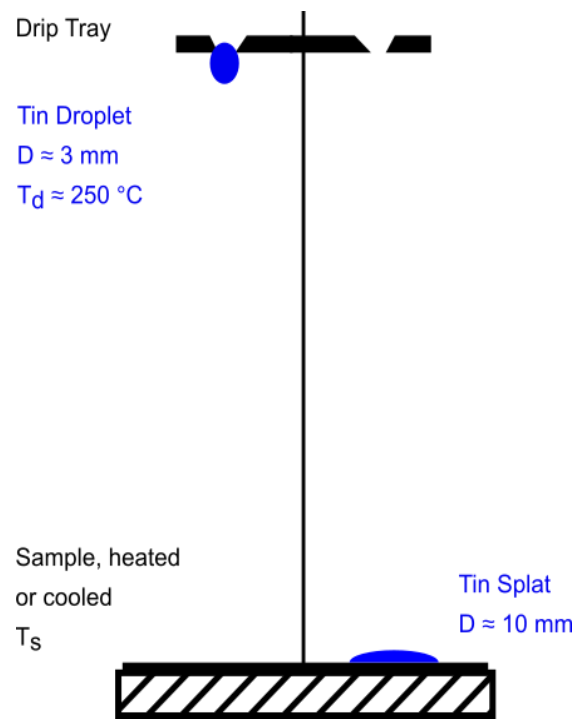

FIG. 2. Sketch of tin dripping scheme.

Samples with up to 6 inch diameter could be mounted with horizontal orientation on a holder made of a copper plate with guard railing at its circumference and connected to its vacuum flange by copper and stainless steel cooling pipes. Figure 3 shows a photo of this sample holder with a mounted silicon wafer with three tin splats on its surface. An aluminum ring and a set of stainless steel clamps were used to press the samples tightly to the holder plate for good thermal contact. Thermocouple wire probes (chromel-alumel type K) were attached to monitor the temperature at several positions on the holder. The holder could be cooled to temperatures down to -150 °C by the flow of cold nitrogen vapor from a cryogenic liquid container with pressure regulation.

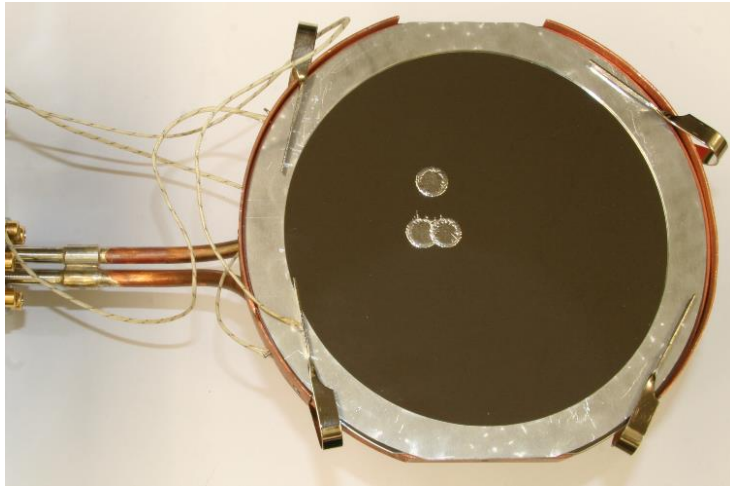

FIG. 3. Photo of sample holder with mounted Si wafer with tin splats.

For initiation of nucleation, tin splats deposited on the substrates could be infected under high vacuum by dropping small seed particles of gray tin powder onto them from above using two small receptacles mounted on an insertable manipulator rod.[23] For phase transformation of tin the temperature $T_s$ of the samples was typically regulated by continuous cooling to the range of -30 to -50 °C. Surface temperatures in the range of between -90 °C and -130 °C were used for tests of in-situ delamination by thermal contraction. Cooling from room temperature at rates of ~7 K/min was employed to reach these low temperatures. To improve the sticking of tin splats to the substrates when dripping, the samples could also be heated up to a temperature of ~90 °C by connecting the pipes of the holder to the flow of hot water vapor generated by boiling of demineralized water.[24] A wire probe mounted on a wobble stick was used to test in-situ the droplet splat adhesion to the respective sample surface. Measurements with a profilometer (Veeco Dektak 3030-ST) were applied to examine the bottom side of delaminated tin splats after removal.

The samples exposed to tin dripping were n-type Si (100) wafers with diameters of 4 and 6 inches. The top surface was super-polished to a roughness of below 0.2 nm root-mean-square. Both uncoated and coated Si wafers (50 bilayers of Mo/Si with $SiO_2$ top layer, optimized for peak reflectance at a wavelength of ~13.5 nm) were used as samples.[27] One 6 inch Si wafer had an etched test grating structure on its surface (straight laminar grating grooves, 0.5 mm wide, ~3 μm deep). In addition, a thick solid silicon carbide substrate (102 mm diameter, 37.5 mm thickness) was examined. In this case, the dripping distance was only 41 cm. The SiC material was produced by a chemical-vapor-composite deposition process (TREX CVC-SiC). Its smooth surface was not super-polished and had a roughness in the μm range. For measurements of the bending stress of droplet splats, tin was also dripped onto very thin (~0.15 mm) cover glass substrates (squares with 18 mm long sides).

For tin removal by snowflake aerosol a hand-held spray gun (Spraying Systems Co.) with expansion of liquid $CO_2$ at a pressure of ~5 MPa through a long filter nozzle (TEM-911-5 filter, 1 mm nozzle diam.) was employed. Expansion from the liquid phase was ensured by use of a gas bottle with riser pipe or by upside-down orientation of the bottle. During expansion a phase change occurs from liquid $CO_2$ to small $CO_2$ snow crystals. A snow jet with a length of about 3 cm and a width of ~1.5 mm was produced with this nozzle, leading to low temperatures at the impact locations on the samples. The samples were scanned with sweeping strokes at some angle of the nozzle during short-term aerosol exposure (several seconds) at a distance of a few cm from nozzle to sample. Industrial grade (99.5% purity) and high-purity grade (99.998%) $CO_2$ was used. For ex-situ tin removal by delamination during thermal contraction, Si wafer samples with sticking tin splats were immersed in an insulated container filled with liquid nitrogen (at T = -196 °C) for a duration of typically 1 minute and then warmed up back to room temperature.

## III. RESULTS
### A. Vacuum characterization

For baseline characterization and comparison we have analyzed the residual gas in the vacuum chamber. It is dominated by water molecules since the system was not baked out. Figure 4 shows a typical residual gas spectrum of the vessel recorded when the total vacuum pressure had reached a fairly low level. In addition to water, major peaks arise from hydrogen (shoulder at scan start), air (attributed to residual virtual leaks) and carbon dioxide. The small peak at m/z = 69 can be ascribed to $CF_3$ fragments arising from fomblin oil used in the turbo-molecular vacuum pump. With respect to its vacuum environment after venting and pump-down our chamber is likely not too different from vessels of commercial EUV sources which are also not baked out.[3,4] Once the $H_2$ background gas is admitted in the scanner source and EUV radiation is generated there it has in comparison a more reductive environment, however. To limit adsorption of water vapor on the sample surface, test runs with cooled sample holder were only carried out when the chamber pressure was in a range below a few $10^{-5}$ Pa.

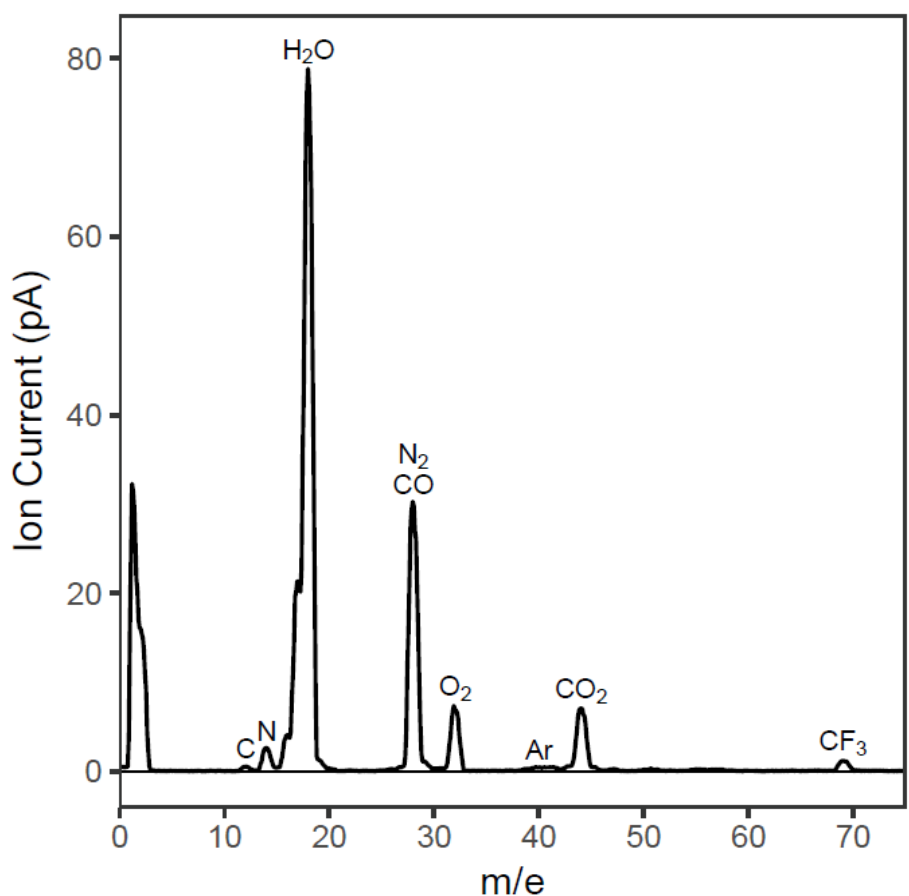

FIG. 4. Residual gas spectrum of the evacuated chamber obtained at a pressure of $1.5 \times 10^{-5}$ Pa. The major gas contributions are identified.

## B.  In-situ phase transformation of tin drops

For smooth 4-inch and 6-inch diameter Si wafers, both with and without Mo/Si multilayer coating, we have carried out dripping of molten tin drops onto the substrates when heated to ~90 °C, leading to adhesion of the splats. The tin deposits were then inoculated with gray tin powder. Next, the phase transformation was carried out during continuous sample cooling at temperatures in the range of -30 °C to -45 °C and monitored by optical imaging with the camera at 5-minute intervals. Regions of gray tin could easily be detected by their reduced light reflection. Typically, full conversion from β-Sn to α-Sn took place within 12 to 17 hours of cooling. It proceeded similarly on coated and uncoated wafers. The transformed tin deposits became brittle and deformed due to expansion. The gray Sn pieces then detached fully from the substrates and moved on the surface due to small vibrations of the sample holder.

The same dripping and transformation sequence was also executed for tin drops on the 6-inch Si grating wafer. When the grating wafer was kept at room temperature during tin dripping the splats did not stick to its surface. However, the tin splats stuck when it was heated to a temperature of around 90 °C during dripping. After room temperature was restored on the sample, the splats were infected with gray tin powder. Figure 5 shows the typical progress of subsequent droplet conversion during cooling to around -40 °C for two adhering tin splats (each with mass ~100 mg) on this wafer in a sequence of five images taken in intervals of 2 hours, and with a final image after 13 hours from the start of cooling. No significant change of splat composition was visible after two hours of cooling, see Fig. 5(b). Then first nucleation sites became visible, clearly seen in the image of Fig. 5(c), taken after 4 hours. The transformation proceeded across the tin splats, Fig. 5(d), with deformation and rupture of expanding gray tin blisters, Fig. 5(e), leading to nearly complete conversion with fragmentation and disintegration of the deposits after 13 hours, see Fig. 5(f).

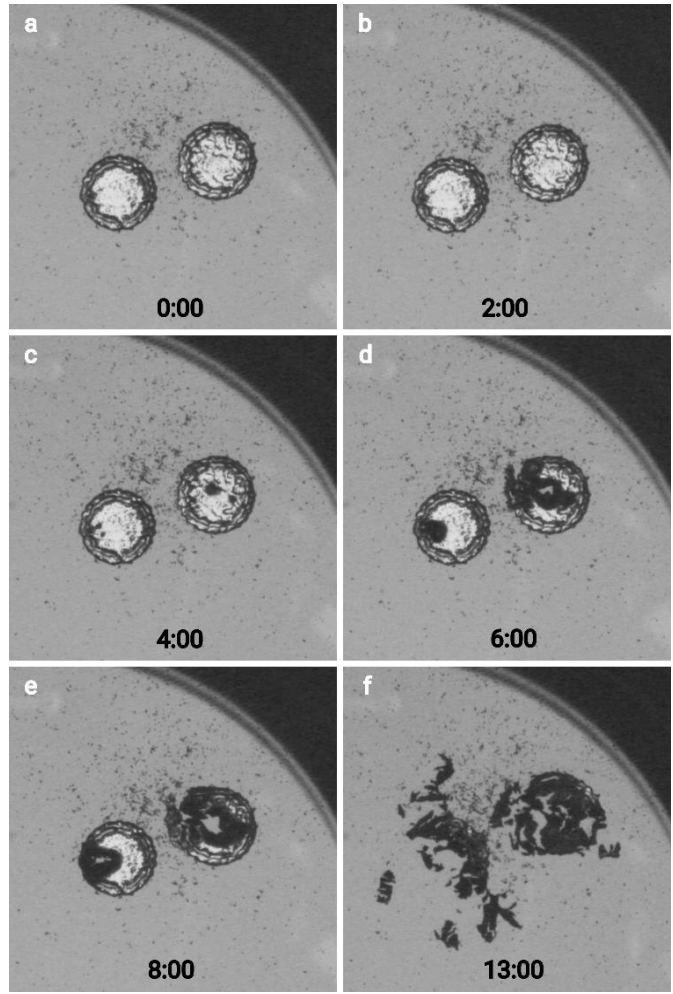

FIG. 5. Images for a section of Si wafer with grating structure recorded during phase transformation of two tin splats: (a) at start of cooling after infection, (b - f) after 2, 4, 6, 8 and 13 hours of cooling, respectively.

Since silicon carbide is a viable candidate for a collector mirror substrate with integrated cooling channels, we have also examined tin dripping and phase transformation on a thick solid piece of SiC. A photo of this sample is shown in Fig. 6 during the mounting process. The substrate surface was heated to ~65 °C during tin dripping. In this case the tin splats did not stick to the sample surface. This behavior is attributed to the roughness of the SiC surface which promotes splat self-peeling even at elevated temperatures.[24] Nevertheless, similar phase conversion of tin drops dripped on the surface could also be achieved within about 10 hours in this case during continuous substrate cooling in the temperature range of -35 °C to -50 °C.

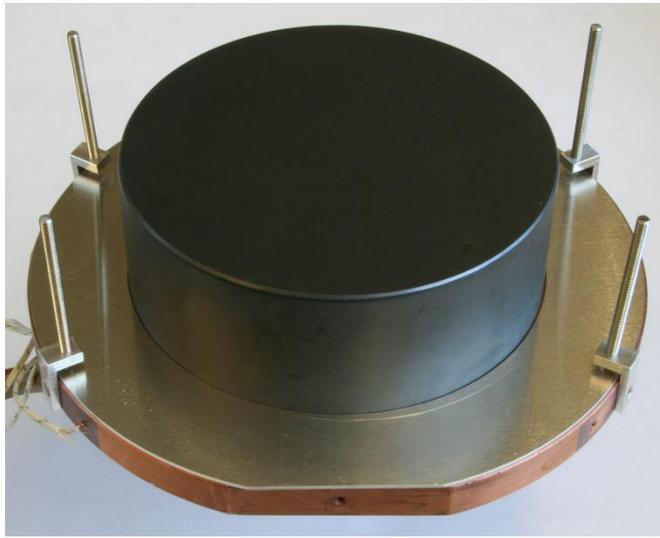

FIG. 6. Photo of silicon carbide substrate on sample holder.

## C.  Tin deposit removal by delamination

For comparison, we have also studied alternative cryogenic techniques for removal of sticking droplet splats. The adhesion of sticking tin splats could not be altered significantly by cooling to -50 °C, corresponding to the conditions used during phase transformation. However, when the sample in the vacuum chamber was more strongly cooled to temperatures of around -120 °C and subsequently warmed up again, we found that the tin droplet splats did no longer stick to the surface. This delamination behavior, attributed to thermal expansion mismatch and contraction strain, was observed for a bare Si wafer, an MLM-coated Si wafer and also for the grating wafer. Figure 7 shows a magnified photo of the bottom side of a tin drop removed by in-situ cooling from the grating wafer. The grating structure became imprinted on the tin splat during its solidification; the corresponding straight grating lines are therefore visible in the photo.

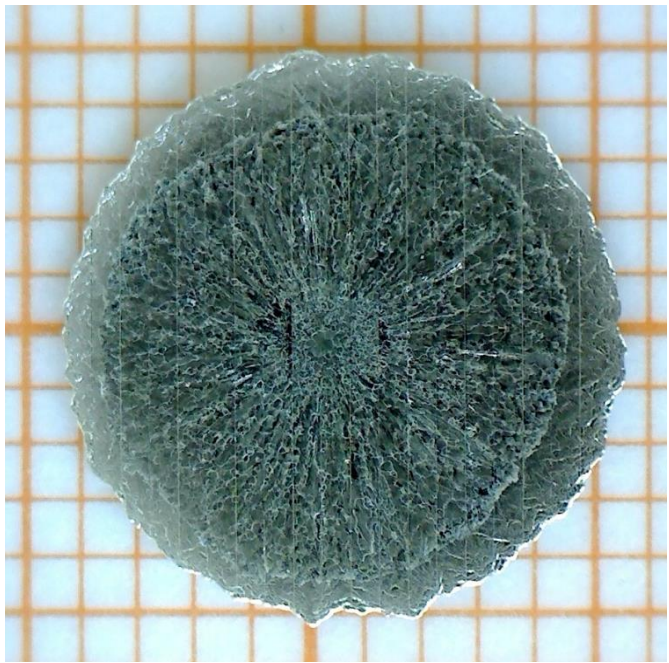

FIG. 7. Magnified photo of tin splat (bottom side) removed from grating sample by strong cooling to -130 °C. Graph paper with mm-size squares is used in the background.

Very similarly, delamination of entire sticking droplet splats was also verified to take place for the case of both coated and uncoated Si wafer substrates by short immersion in liquid nitrogen in an ex-situ cryogenic process. Here, the detachment of sticking splats was generally observed already during the rapid cooling and before the substrate was warmed up back to room temperature. In contrast, however, for tin drops deposited onto a smooth glass slide at $T_s = 94$ °C, removal was not possible by this immersion technique. Due to the very strong bonding of tin splats to the glass surface in this case, the only successful cryogenic removal technique surmounting the splat adhesion was induction of tin phase transformation.

Using the profilometer, we have scanned the bottom side of the tin drop that was removed by immersion in liquid nitrogen from the grating wafer. A diagonal scan across the splat in a direction perpendicular to the grating grooves is shown in Fig. 8. The imprinted grating structure and the curvature of the splat due to bending stress created by surface tension are visible. For the upper curve (a) a spline fit was applied to reproduce the curvature and subtract it from the data. The lower plot (b) shows a comparison of the subtraction result with a direct scan of a section of the grating wafer surface. The close matching of the traces illustrates that the grating structure is indeed reproduced fairly closely by the tin splat during solidification after droplet impact and preserved after splat peeling during thermal contraction.

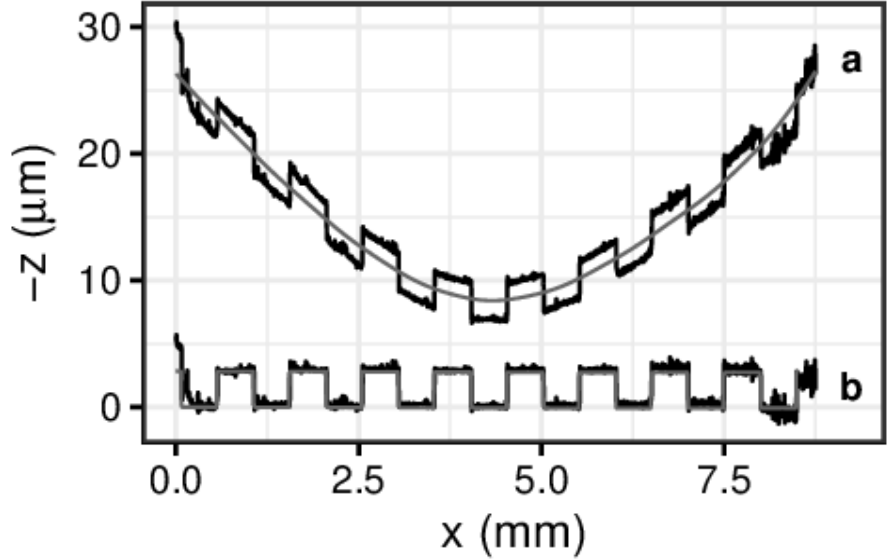

FIG. 8. Profilometer scans: (a) scan across bottom surface of delaminated tin drop and spline fit to curvature, (b) data of bottom surface with curvature subtracted in comparison to a scan of a section of grating substrate.

For comparative measurements of the induced bending stress we have dripped molten tin in air onto the thin cover glass substrates. By use of the profilometer the backside surface of the glass piece was scanned before and after tin dripping in order to detect the changes induced due to the stress of adhering deposits. Underneath tin splats with diameters of ~10 mm, the magnitude of bending excursions was found to be typically in the range of a few µm.

### D. Tin splat removal by snowflake jetting

As a further alternative cryogenic removal technique for sticking tin deposits we have also examined the impact of $CO_2$ snowflake aerosol jets. Using the hand-held spray gun, this method was applied ex-situ under atmospheric conditions. The photos of Fig. 9 were taken during and after snow-jet cleaning of tin drops on a MLM-coated Si wafer with 4 inch diameter. By sweeping the aerosol jet across the sample surface the adhering tin splats and deposits could be removed within just a few seconds of exposure. However, when the lower-purity grade $CO_2$ gas was used (as in the case of Fig. 9) a very thin residue became visible on the sample surface after prolonged application, see Fig. 9b. As already previously described and investigated by Zito[12], liquid carbon dioxide is a good solvent for oils and other organic impurities. Therefore, when industrial grade $CO_2$ gas is used for cleaning, organic contaminants may be deposited. On the other hand, surface contamination is avoided when using spectrally pure gas, as was confirmed in our tests of tin splat removal when using the carbon dioxide gas with high purity. No residue was observed in this case.

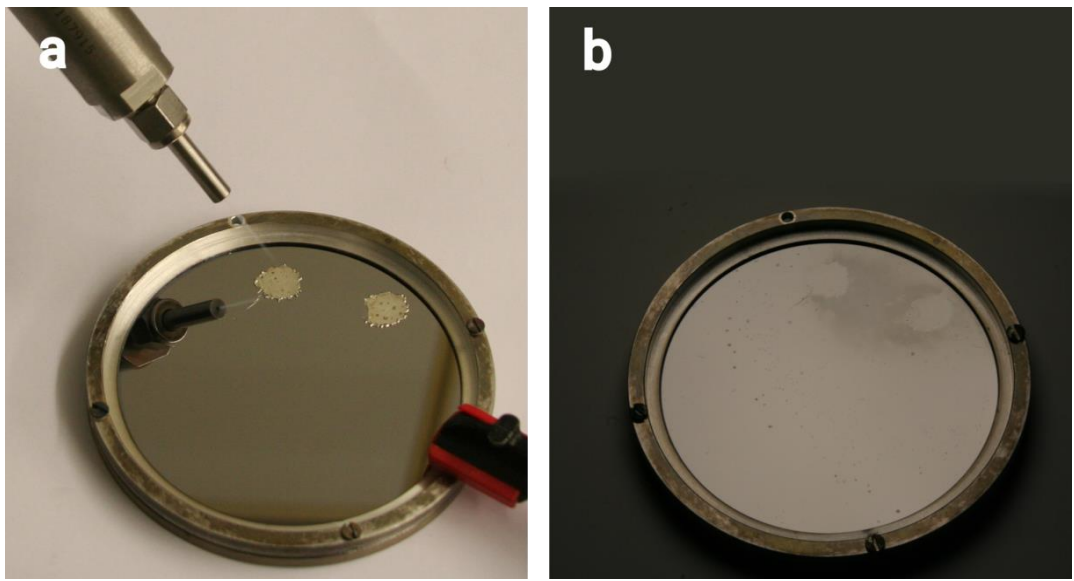

FIG. 9. (a) Photo of setup for $CO_2$ snow jet cleaning showing spray nozzle and tin droplet splats on 4-inch MLM-coated Si wafer mounted in a holder. (b) Photo of the same wafer after removal of the tin droplets. The photo was taken under special illumination conditions to make the thin layer of residue near the tin splat locations visible.

## IV. DISCUSSION

Heating of the examined substrates resulted in higher contact temperatures at the smooth surface during tin dripping, leading to stronger sticking of solidifying droplet splats due to increased adhesion.[24,28] Strong sticking was also found when tin drops were incident on already existing splats on the surface. In the case of EUV sources a large heat flux is originating from the hot plasma. Similar contamination effects by tin drops were observed on the collector mirrors.[5] The surface temperature of the mirror is raised during source operation and severe contamination with adhesive tin splats can occur even when cooling is applied to the mirror. A tin cleaning technique that can be carried out in-situ represents an attractive solution for this problem.

During our phase-conversion experiments the visible onset of α-Sn nucleation on tin deposits showed some time variation and occurred only in the vicinity of a few seed crystals, confirming the statistical nature of this initiation process. However, when many small seeds were placed on the splats the phase conversion began always at some locations on the tin surface within a few hours after start of cooling. The subsequent spread of gray tin regions on the surface and into the bulk, on the other hand, proceeded fairly rapidly at all nucleation sites, with growth rates of 10–15 μm/min, leading consistently to nearly complete tin transformation of contiguous splats in less than 24 hours. After the samples were taken out of the chamber, inspection showed that the contamination was indeed fully removed without any residues. It was shown previously[20] that such fast transformation rates to α-Sn can be achieved provided the material purity grade of tin is 99.999%. The required purity for operation of the tin droplet generator in the commercial source is also at a level of at least 99,999%.[29] Since it is unlikely that the tin in the drops will be contaminated at the chamber walls or at other places with elements that are known to hinder the transformation[19,20] it can be safely assumed that rapid tin transformation will also be reached during cooling of infected tin splats sticking to the surface of the collector mirror in the EUV source of the scanner.

The successful demonstration of in-situ removal of adhering tin deposits on large samples by phase transition confirm our earlier results that were obtained for small samples.[24] During the phase transformation the bonding character changes from metallic in the case of the β-Sn phase to covalent, with sp$^3$ hybrid bonds, for the semiconducting phase α-Sn. The volume of tin increases substantially (by around 26%) during the rearrangement of bonds in the lattice and the resulting gray tin crystals become very brittle. This explains the strong reduction of adhesion. The bonding to the surface is strongly weakened because macroscopic movements of the adhering deposits occur during the crystal transformation leading to the break-up of bonds with the surface and also to lateral displacements relative to the substrate surface. Several decades ago, adhesion and friction behavior of group IV elements at a gold surface was compared in experiments by Buckley.[30] The results indicated that the more covalently bonded lighter elements (Si and Ge) exhibited lower adhesion and friction compared to the heavier group IV elements with metallic bonds (Sn and Pb), thus confirming the influence of the bonding character on the adhesion strength.

Our results show that sticking tin splats could be removed from large samples by phase conversion, independent of whether a coating with Mo/Si was present or not on the sample, whether a grating structure was etched on the surface or not, whether the sample was thin or thick. Even though the thermal transport from the solidified tin droplet to the bulk of the substrate is slightly impeded in the case of a coating with MLM as compared to bare Si wafers,[31] the tin splat sticking and cooling behavior can still be considered to be fairly similar for these sample types. Since the thermal properties describing the substrate effusivity $e_i$[24] are not very different for crystalline Si and CVC-SiC ($e_{Si}$ = 15675 $Ws^{1/2}m^{-2}K^{-1}$, $e_{SiC}$ = 19880 $Ws^{1/2}m^{-2}K^{-1}$), a quite similar transformation behavior may be expected for these substrate materials as well (for similar surface roughness) at the interface region with tin deposits. In the case of the grating sample, examination of the bottom surface of delaminated tin splats showed that the grating structure was fully replicated by the molten tin during its solidification on the surface (see Figs. 7, 8). Thus, a stronger adhesion force may be surmised in this case. Nevertheless, the tin splats could be fully removed, since a complete embrittlement of the deposits could be achieved by the phase transformation.

For splat removal by delamination during strong cooling, on the other hand, the crystal phase does not change and the tin metal keeps its ductile properties. However, in this case differences in contraction between sticking tin deposits and substrates can influence the adhesion due to thermally induced interfacial shear stress. Table 1 lists the coefficients of thermal expansion (CTE) for the relevant materials at room temperature. According to these data, a temperature change from room temperature by 150 K or 220 K would lead to an estimated dimensional change of 0.33 % or 0.48 %, respectively, in case of a solid piece of β-Sn, whereas it would be more than 8 times less in case of the Si surface. The corresponding strain by expansion mismatch can induce a sufficiently large local displacement on the sample surface to initiate complete delamination of tin splats with diameters of several mm, since the resulting surface tension cannot be fully compensated by the force of adhesion during thermal contraction and expansion. For the multilayer coating, on the other hand, the expansion mismatch to the Si substrate is much smaller during cooling. We have verified that no coating delamination occurred even after prolonged and cyclic sudden immersion of MLM samples in liquid nitrogen.

TABLE I. Coefficient of thermal expansion at T = 300 K for β-Sn, Si and SiC.

| Material | CTE |
|---|---|
| β-Sn | $2.2 \times 10^{-5}$ K$^{-1}$ |
| Si | $2.6 \times 10^{-6}$ K$^{-1}$ |
| SiC | $2.3 \times 10^{-6}$ K$^{-1}$ |

At the bottom side of delaminated or non-sticking droplets we generally observe imprinted radial flow patterns pointing from the center of the tin splat to its circumference (see figure 7), in agreement with earlier findings.[24,32] This indicates that solidification occurs very fast during the spreading of the tin drops on the substrates which happens in just a few milliseconds after central impact.[28,32] Nevertheless, the initially liquid tin fully replicates the grating structure, as demonstrated by the profilometric analysis shown in figure 8. As discussed previously, thermally induced bending stress builds up during solidification of the droplets.[24,28] Therefore, adhesion is less strong at the edge of the splat where it tends to bend away from the sample surface. As a result, after delamination the bottom sides of tin splats show a slight curvature on a scale of micrometers and are not completely flat, as seen for example in the data of Fig. 8 (a). Correspondingly, tensile stress was also detected by the measured bending curvature during scans of the backside of the thin glass plates underneath the tin droplet splats. Typically, the stress was determined to be in the range of a few MPa.

For snowflake jet cleaning the temperature change is not as large, in comparison to delamination by contraction. However, the $CO_2$ clusters formed by sudden expansion enable in addition a momentum transfer to the tin deposits that also promotes contamination removal. Once the tin splat is slightly lifted at its edge the impact thrust of the snow jet peels it off from the sample surface altogether quite readily, exceeding its adhesion. This technique provides a dry, chemically inert and in principle residue-free removal process when high-purity gas is used. Other studies in connection with EUV mask cleaning have already shown that there is no negative impact on the sample surface or EUV reflectance of the MLM with this method.[13] However, when exposures with durations longer than a few seconds are required, moisture condensation has to be avoided in order to exclude the humidity-induced accumulation of small water droplets on the sample surface. In addition, the removal process could be improved by using an automated mechanical mount for repeated sweeping strokes across the sample surface. We have applied the hand-held technique for comparison with the other cryogenic cleaning schemes and did not try to optimize this arrangement any further.

Based on our results with respect to the cleaning of tin drops it may be concluded that stronger cooling may be beneficial for the operation of a collector mirror with internal cooling channels in a scanner source. However, many different aspects have to be taken into account during the operation of an LPP source. For example, in addition to the sticking of incident Sn particles and droplets, which decreases with decreasing temperature,[24] the surface temperature will also affect the efficiency of continuous in-situ etching and cleaning by hydrogen radicals that takes place during source operation[4,7]. If cooling to negative Celsius temperatures is desired the cooling water circuit for the collector mirror could be replaced by a circuit using a cooling fluid based on ethylene glycol in order to reach lower temperatures. A corresponding temperature management system could then enable in-situ allotropic tin transformation of adhesive tin splats during off-times when the source is not generating any EUV light. For flexibility and convenience we have used in our studies cold nitrogen vapor to provide the cooling. However, in a commercial source the use of an anti-freeze fluid for cooling would be more suitable.

Contaminations on EUV collector mirrors with internal cooling capability could be cleaned inside of commercial source vessels by tin phase conversion if efficient inoculation can be produced to start the process. One possibility would be the inflow of gray tin nanoparticles imbedded in a small gas stream. However, this may not necessarily be required. Light sources are usually operated in a low-pressure hydrogen background gas environment and the EUV radiation present leads to efficient production of hydrogen radicals.[3,4] Since the formation of gray tin nanocrystals by chemical sputtering of β-Sn targets was experimentally observed in hydrogen plasmas even at temperatures above room temperature (up to 60 °C),[33] it may be anticipated that the required α-Sn seed particles may be produced and already be available in sufficient quantity during normal EUV source operation. The interaction of hydrogen radicals present in the source environment may also promote directly the start of phase transformation of tin deposits following diffusion that could lead to a reduced strength of the β-Sn matrix by generating increased tensile stress.[19]

## V.   SUMMARY AND CONCLUSIONS

For the removal of sticking tin splats from large test substrates, we have compared the phase-transformation technique with other cryogenic cleaning schemes, based on thermal delamination initiation or aerosol snowflake impact. All employed methods are fairly simple and have low costs.  They resulted in fast and efficient removal of tin deposits. Embrittlement of tin contamination by phase transformation could be achieved inside of a vacuum chamber not only for coated and uncoated substrates with smooth surfaces but also for a grating-structured sample with sticking tin splats and for a SiC substrate. Likewise, strong cooling with nitrogen vapor enabled in-situ delamination of contiguous tin deposits on smooth wafer surfaces. Alternatively, ex-situ cryogenic removal was examined successfully for $CO_2$ snow-jet impact and for sample immersion in liquid nitrogen. The examined cryogenic cleaning techniques are suitable for application to MLM EUV optics, but they could also be applied to remove thick tin contaminations from chamber walls at other locations if cooling could be provided. Compared to methods of tin-etching by hydrogen radicals, all cryogenic techniques examined here are faster by orders of magnitude. Therefore, the application of such cleaning techniques could be useful for commercial EUV sources.

The location and orientation of the collector mirror at the bottom of the source chamber in the current generation of EUV scanners[4] is certainly a disadvantage with respect to the accumulation of tin contamination on its surface. In contrast, the design of next-generation EUV scanners with high numerical aperture provides a more favorable collector mirror position due to horizontal source orientation[34] so that tin pieces converted in-situ could slide down from the collector mirror according to the force of gravity after their embrittlement and get collected. In-situ tin cleaning could extend the mirror lifetime between collector swaps. In comparison, application of ex-situ collector cleaning concepts appears to be less attractive, since after the exchange of a collector module sufficiently low-pressure vacuum conditions first have to be established again by corresponding time-consuming pumping before light source operation can be resumed.

## ACKNOWLEDGEMENTS

Our study was motivated by the ongoing industrial development of EUV light sources. We are grateful to the molecular and surface physics group at Bielefeld University for general support and for supplying Mo/Si-multilayer-coated EUV mirror samples. Furthermore, we would like to thank Torsten Feigl and his team at optiX fab for generously providing a silicon wafer sample with etched grating structure for our tests. Sincere thanks also go to Kyle Webb for arranging for a substrate sample made of SiC and to Kay Hoffmann for continuous encouragement. We also thank Mark van de Kerkhof for discussions of the ASML scanner-source environment. This research did not receive any specific grant from funding sources in the public, commercial, or not-for-profit sectors.

**References:**

[1] A. Yen, H. Meiling, and J. Benschop, IEEE International Electron Devices Meeting (IEDM) 11.6.1 (2018).

[2] I. Fomenkov, A. Schafgans, and D. Brandt, Synchrotron Radiation News **32,** 3 (2019).

[3] H. Mizoguchi *et al.*, Proc. SPIE **10583,** 1058318 (2018).

[4] I. Fomenkov *et al.*, Adv. Opt. Technol. **6,** 173 (2017).

[5] Y. C. Chen, S. K. Yu, C. Yang, S. C. Chien, L. J. Chen, and P. C. Cheng, 2020 U. S. Patent Application 2020/0004167 A1 (2 January 2020).

[6] W. A. Soer, M. M. J. W. van Herpen, M. J. Jak, P. Gawlitza, S. Braun, N. N. Salashchenko, N. I. Chkhalo, and V. Y. Banine, J. Micro-Nanolith. MEM **11,** 021118 (2012).

[7] D. C. Brandt *et al.*, Proc. SPIE **9048,** 90480C (2014).

[8] D. Elg, J. R. Sporre, G. A. Panici, S. N. Srivastava, and D. N. Ruzic, J. Vac. Sci. Technol., A **34** 021305 (2016).

[9] D. Elg, G. A. Panici, S. Liu, G. Girolami, S. N. Srivastava, and D. N. Ruzic, Plasma Chem. Plasma **223,** 38 (2018).

[10] J. Baek, M. C. Abraham, D. R. Evans, and J. M. Gazza, U. S. Patent No. 9,888,554 (6 February 2018).

[11] G. A. Panici, D. Qerimi, and D. N. Ruzic Proc. SPIE, **10957,** 109571B (2019).

[12] R. R. Zito, Proc. SPIE **1236**, 952 (1990).

[13] I. Varghese, C. W. Bowers, and M. Balooch, Proc SPIE **8166**, 816615 (2011).

[14]M. Moriya, Y. Ueno, T. Abe, and A. Sumitani, U. S. Patent No. 8,256,441 (4 September 2012).

[15]M. Becker, U. Müller and O. Arp, PCT Patent Application WO 2015/043833 A1 (2 April 2015).

[16]A. S. Kuznetsov, K. Bystrov, V. Y. Banine, M. A. van de Kerkhof, N. Schuh, and A. Nikipelov, PCT Patent Application WO 2019/091708 A1 (16 May 2019).

[17]A. De Dea, M. Varga, A. I. Ershov and R. L. Morse, U. S. Patent No. 9,073,098 (7 July 2015).

[18]T. Feigl *et al.*, Proc. SPIE **8679**, 86790C (2013).

[19]W. J. Plumbridge, J. Mater. Sci: Mater. Electron. **18**, 307(2007).

[20]N. Böwering, Mater. Chem. Phys. **198**, 236 (2017).

[21]M. Machida, U. S. Patent Application 2016/0062251 A1 (3 March 2016).

[22]E. Hosler, U. S. Patent Application 2017/0252785 A1 (7 September 2017).

[23]N. Böwering and C. Meier, J. Vac. Sci. Technol. B **36**, 021602 (2018).

[24]N. Böwering and C. Meier, Appl. Phys. A **125**, 633 (2019).

[25]M. Kriese *et al.*, Proc. SPIE **9048,** 90483C (2014).

[26]T. Feigl *et al.*, Proc. SPIE **9422**, 94220E (2015).

[27]U. Kleineberg, T. Westerwalbesloh, W. Hachmann, U. Heinzmann, J. Tümmler, F. Scholze, G. Ulm, and S. Muellender, Thin Solid Films **433**, 230 (2003).

[28]J. De Ruiter, D. Soto, and K. K. Varanasi, Nat. Phys. **14**, 35 (2018).

[29]G. Vaschenko, P. Baumgart, C. Rajyaguru, B. Sams, A. Ridiger, and J. Kardokus, U. S. Patent No. 10,455,680 (22 October 2019).

[30]D. H. Buckley, NASA Technical Note, **TN D-7930**, (1975).

[31]E. Bozorg-Grayeli, Z. Li, M. Asheghi, G. Delgado, A. Pokrovsky, M. Panzer, D. Wack, and K. E. Goodson, J. Appl. Phys. **112**, 083504 (2012).

[32]S. Shakeri and S. Chandra, Int. J. Heat Mass Transf. **45**, 4561 (2002).

[33]H. Fukumoto, H. Myoren, K. Matsumoto, T. Imura, Y. Osaka, and M. Ichihara, Solid State Commun. **62**, 309 (1987).

[34]A. Yen, Presentation at Semicon West Conf. (2019).